# Study of formation of stable fragments in central heavy-ion collisions.

Supriya Goyal and Rajeev K. Puri*
*Department of Physics, Panjab University, Chandigarh-160014, INDIA*
*\* email: rkpuri@pu.ac.in*

## Introduction

The multifragmentation has been of central interest in last two decades. The primary cause of this intense investigation is the rich information associated with the production and emission of fragments [1,2]. A large number of experimental and theoretical attempts are reported in the literature [1,2]. On the theoretical front, in the mean field approaches, light ion reactions and production of light complex clusters can be handled via two and three body problems. Alternately, one can employ phase space models such as Quantum Molecular Dynamics (QMD) model [1] and Boltzmann-Uehling-Uhlenbeck (BUU), to follow the reaction from the start where colliding nuclei are well separated to the end where matter is cold and fragmented. Since no phase space model simulates the clusters, therefore, we need to apply after burners or secondary algorithms to form clusters [2]. In the past, one has devised spatial correlation procedure to clusterize the phase space i.e. Minimum Spanning Tree (MST) method [2]. The problem with this algorithm is that the clusters, especially the light clusters are not properly bound. Several improvements were suggested in this direction [3]. All of these approaches demand binding energy to be negative with a value varying between 0 and 4 MeV/nucleon. In the present study, we extend MST method by implementing binding energy at a microscopic level (named as MSTB (2.1)) i.e. each fragment is subjected to its true binding energy calculated using the modified Bethe-Weizsäcker (BWM) formula [4]. The nucleons of a cluster that fails to fulfil the BWM binding energy criteria are treated as free nucleons. The model used for the present study is Quantum Molecular Dynamics (QMD) model.

## Model

The QMD model is an n-body theory that simulates the heavy-ion reactions on event by event basis [1]. This is based on a molecular dynamic picture where nucleons interact via two and three-body interactions. The nucleons propagate according to the classical equations of motion:

$$\frac{d\mathbf{p}_i}{dt} = -\frac{dH}{d\mathbf{r}_i} \quad \text{and} \quad \frac{d\mathbf{r}_i}{dt} = \frac{dH}{d\mathbf{p}_i}, \quad (1)$$

where H stands for the Hamiltonian which is given by

$$H = \sum_i \left[ T_i + \frac{1}{2} \sum_{ij} V_{ij} \right]. \quad (2)$$

$T_i$ is the kinetic energy term and $V_{ij}$ is the nuclear potential which consists of

$$V_{ij} = V^{Skyrme} + V^{Yuk} + V^{Coul}, \quad (3)$$

where $V^{Skyrme}$, $V^{Yuk}$ and $V^{Coul}$ are, respectively, the local (two and three-body) Skyrme, Yukawa and Coulomb potentials. The modified Bethe-Weizsäcker formula (BWM) is given as

$$E_{bind}^{BWM} = a_v N_f - a_s N_f^{2/3} - a_c \frac{N_f^z (N_f^z - 1)}{N_f^{1/3}} - a_{sym} \frac{(N_f - 2N_f^z)^2}{N_f \left(1 + e^{-N_f/17}\right)} + \delta_{new}. \quad (4)$$

The pairing term $\delta_{new}$ is given by:

$$\delta_{new} = +a_p N_f^{-1/2} \left(1 - e^{-N_f/30}\right) \text{ even } N_f^z \& N_f^n \quad (5)$$

$$\delta_{new} = -a_p N_f^{-1/2} \left(1 - e^{-N_f/30}\right) \text{ odd } N_f^z \& N_f^n \quad (6)$$

$$\delta_{new} = 0 \quad \text{odd } N_f \quad (7)$$

The strength of various parameters is listed in Ref. [4].

## Results and discussion

Here we use a soft equation of state along with energy dependent nucleon-nucleon cross-section. We concentrated on the symmetric as well as asymmetric reactions, as the response of the same model can be quite different for both types of reactions. It is shown in Ref. [5] that MST can explain the multiplicity of fragments in symmetric or nearly symmetric reactions nicely whereas it fails badly to predict the fragments in highly asymmetric reactions. Since the new modification is meaningful only for highly excited systems, therefore, we considered central collisions only. The reactions are followed till saturation time which, in the present work is between 200-300 fm/c and then fragments are identified. In fig. 1, we display the final state spatial coordinates of nucleons/fragments (x-z plane) for the reactions of $^{197}$Au+$^{197}$Au at b=5fm and E=600MeV/nucleon and central reactions of $^{16}$O+$^{80}$Br and $^{16}$O+$^{108}$Ag at incident energy of 200MeV/nucleon. The left panel is using the MST method whereas right panel indicates clusters after binding energy filter i.e. after MSTB (2.1) method is implemented. We see that significant number of fragments fails to fulfil the binding energy check and hence are eliminated in the MSTB(2.1) procedure. This situation is most grave in the reaction of O+Br/Ag where even the heaviest fragment fails to fulfil the binding energy check. Obviously, unstable fragments have large number of nucleons with large relative momenta [6].

## Acknowledgments

This work is supported by grant from Department of Science and Technology (DST), Govt. of India.

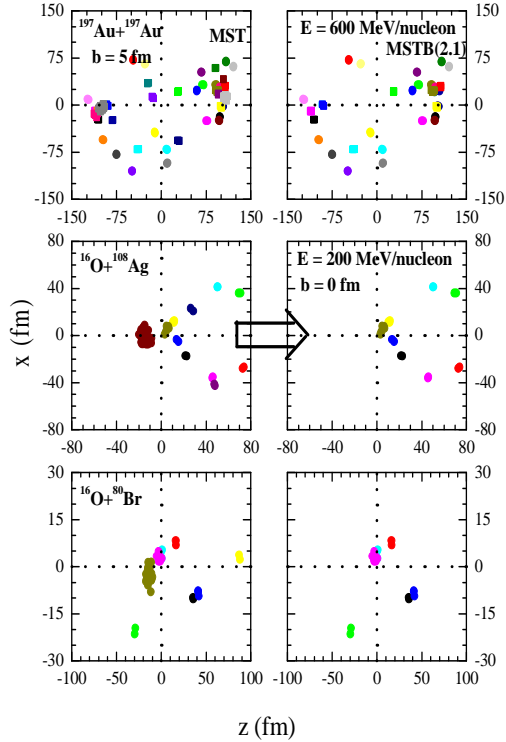

**Fig. 1** The snap-shot of spatial coordinates of nucleons in the (x-z) plane. Details are given in the text.